\documentclass[aps,prl,reprint,superscriptaddress,nofootinbib]{revtex4-2}

\usepackage{amsmath,amssymb,bm}
\usepackage{graphicx}
\usepackage{booktabs}
\usepackage{array}
\usepackage{longtable}
\usepackage{float}
\usepackage{xcolor}
\usepackage[hidelinks]{hyperref}
\usepackage{microtype}
\usepackage[normalem]{ulem}

\newcommand{\DSP}{D_{SP}}
\newcommand{\RphiBV}{R_{\phi/BV}}

\newcommand{\cc}{\mathrm{c.c.}}

\begin{document}

\title{Decipher the nature of glueball candidate $X(2370)$}

\author{Sheng-Qi Zhang}
\affiliation{Center for High Energy Physics, Peking University, Beijing 100871, China}

\author{Bing-Dong Wan}
\email{wanbd@lnnu.edu.cn}
\affiliation{School of Physics and Electronic Technology, Liaoning Normal University, Dalian 116029, China}
\affiliation{Center for Theoretical and Experimental High Energy Physics, Liaoning Normal University, Dalian 116029, China}

\author{Cong-Feng Qiao}
\email{qiaocf@ucas.ac.cn}
\affiliation{School of Physical Sciences, University of Chinese Academy of Sciences, Beijing 100049, China}

\begin{abstract}
As a unique form of matter composed entirely of gauge bosons, glueballs are an important low-energy prediction of QCD. After decades of searches, the BESIII Collaboration recently suggested that $X(2370)$ may contain a dominant glueball component, based largely on evidence that it is approximately a flavor singlet. Recognizing that flavor-singlet character does not uniquely identify a glueball, we perform a comprehensive analysis using available mass, flavor-singlet, and decay constraints. We find that three flavor-singlet configurations—hybrid meson, tetraquark state, and trigluon glueball—can satisfactorily reproduce the existing experimental data. Among these possibilities, the hybrid structure best describes the measured three-pseudoscalar decay ratios. To ultimately pin down the dominant structure of $X(2370)$, we propose measuring the decay-width ratios $a_0(1450)\pi/[K_0^*(1430)\bar K+\mathrm{c.c.}]$ and $\phi\phi/[b_1(1235)\rho]$, which can exclusively distinguish these three scenarios.
\end{abstract}

\maketitle

Glueballs are color-singlet hadrons generated by gluonic self-interactions, and their identification would provide direct spectroscopy of the non-Abelian gauge sector of QCD. Lattice QCD calculations place the lightest pseudoscalar glueball in the mass range $2.3$--$3.0$ GeV~\cite{Bali:1993fb,Morningstar:1999rf,Chen:2005mg,Gregory:2012hu,Gui:2019dtm,Vadacchino:2023vnc}.  Within this range, $X(2370)$ was observed in gluon-rich radiative $J/\psi$ decays with $M_X=2359^{+13}_{-14}$ MeV and $J^{PC}=0^{-+}$~\cite{BESIII:2010gmv,BESIII:2023wfi,BESIII:2026rzt}, making it a leading pseudoscalar glueball candidate.

More recently, BESIII observed a strong suppression of $X(2370)\to K^*\bar K+\cc$ and identified it as evidence that $X(2370)$ is approximately a flavor singlet~\cite{BESIII:2026mvn,LIPKIN1981114,LIPKIN1982326}. This result provides additional support for the glueball interpretation of $X(2370)$ because flavor singletness is an essential property of a glueball~\cite{BESIII:2026mvn}.

Nevertheless, this evidence constrains flavor more directly than constituent structure. The generalized $G$-parity selection rule underlying the $K^*\bar K$ suppression probes flavor quantum numbers~\cite{LIPKIN1981114,LIPKIN1982326} and can be satisfied by any flavor-singlet $0^{-+}$ configuration, including $q\bar q$, hybrid, multiquark, and glueball states. Mass compatibility, radiative production, and decay patterns are likewise shared by distinct constituent assignments~\cite{Yu:2011ta,Wang:2017iai,Sun:2021kka,Dong:2020okt,Su:2022eun,Wang:2025nme,Lu:2026foy,Zhang:2022obn,Wan:2021vny,Wang:2026iyq,Eshraim:2012kr,Giacosa:2023lmi,Giacosa:2024sgv}. The central question is therefore which flavor-singlet structure dominates $X(2370)$ and how the alternatives can be distinguished experimentally.

We consider seven representative assignments: a conventional $q\bar q$ state, a flavor-singlet $q\bar qg$ hybrid, a compact $P$-wave $qq\bar q\bar q$ tetraquark, a compact hexaquark, $\Lambda\bar\Lambda$ baryonium, a two-gluon glueball ($2g$), and a trigluon glueball ($3g$). Here the $2g$ and $3g$ labels denote dynamic gluon assignments in constituent descriptions. Full-QCD gluon self-interactions mix Fock sectors with different gluon numbers and preclude a unique fixed-gluon-number assignment, while the associated color--spin structures can still generate distinct decay patterns. These assignments are assessed against existing mass, flavor, and decay constraints, with further discrimination sought through decay-width ratios. All details of the calculations can be found in the Supplemental Material. 


An ordinary radial $\eta/\eta'$-like assignment is disfavored by the combined experimental constraints. BESIII obtains $R_{K^*\bar K}<0.081$ and infers ${\cal B}(X\to K^*\bar K)<1.6\%$ and a corresponding partial width below about $2$ MeV~\cite{BESIII:2026mvn}. Conventional radial-pseudoscalar calculations instead give $\Gamma(X\to K^*\bar K)\simeq15$--$200$ MeV, while a phase-space extrapolation from $\eta(1405)/\eta(1475)$ also gives a width well above $20$ MeV~\cite{Yu:2011ta,Wang:2017iai}. The simultaneous suppression of $X\to\gamma\omega$ and $\gamma\phi$, together with the absence of an additional flavor-singlet light-quark level in lattice spectroscopy near this region~\cite{BESIII:2026mvn,Dudek:2011tt}, reinforces this conclusion. We therefore do not retain an ordinary light-quark-dominated $q\bar q$ pseudoscalar as an independent pure-state candidate.

The two specific hexaquark assignments are next tested against the relevant mass, flavor, and decay constraints. For $\Lambda\bar\Lambda$ baryonium, an SU(3)-broken eight-dimensional chromomagnetic calculation gives only a $\Lambda\bar\Lambda$ probability $P_{\Lambda\bar\Lambda}\simeq3.5\%$ for the eigenstate closest to $X(2370)$, whereas the $\Lambda\bar\Lambda$-dominated state lies near $2.26$ GeV. Thus the state compatible with the measured mass is not dominated by the physical $\Lambda\bar\Lambda$ component. Independently, a quasipotential Bethe--Salpeter (qBSE) analysis continued to the second Riemann sheet finds that reproducing the measured pole requires an emissive rather than physically absorptive optical interaction, showing that the $\Lambda\bar\Lambda$ interaction does not dynamically generate $X(2370)$. The absence of both a $\Lambda\bar\Lambda$-dominated eigenstate near the measured mass and a physically acceptable pole therefore rules out the predominantly $\Lambda\bar\Lambda$ baryonium assignment within the frameworks tested here.

We next test whether a single compact-hexaquark eigenstate can simultaneously satisfy the mass, flavor, and three-pseudoscalar decay constraints. Before flavor-changing annihilation is included, a Pauli-complete 32-dimensional calculation produces a level near $2.38$--$2.39$ GeV, but its eigenvector has a flavor-octet probability $P_8\simeq60\%$ and a flavor-singlet probability $P_1\simeq37.5\%$. Once the annihilation interaction is fixed independently from the $\eta$--$\eta'$ spectrum, the singlet-dominated trajectory moves to $2.61$--$2.63$ GeV, while the state remaining near $X(2370)$ gives $\chi^2_{4PPP}\simeq26$--$29$ for the four measured $PPP$ ratios. This simultaneous mismatch disfavors the specific compact-hexaquark assignment considered here. 


The remaining two-gluon glueball, trigluon glueball, flavor-singlet-hybrid, and tetraquark alternatives are compared through their three-pseudoscalar decay patterns. We define the relative widths
\begin{equation}
R_f^{PPP}=\frac{\Gamma(X\to f)}{\Gamma(X\to K\bar K\pi)},
\qquad R_{K\bar K\pi}^{PPP}=1.
\label{eq:ppp_intro}
\end{equation}

\begin{table*}[t]
\caption{One-parameter fits of the calculated $PPP$ decay patterns to the product branching fractions in $J/\psi\to\gamma X\to\gamma PPP$. The four rightmost columns give the signed pulls in the indicated channels; boldface marks the smallest absolute pull in each channel. The tetraquark entry is obtained from an interpolating-current calculation for the physical compact $P$-wave tetraquark considered below. Experimental correlations and a common theory covariance are unavailable; the quoted $p$ values therefore characterize compatibility only within the stated error treatment.}
\label{tab:pppfit}
\centering
\small
\setlength{\tabcolsep}{5pt}
\begin{tabular}{lccccccc}
\toprule
Decay pattern & $A_{3P}\ (10^{-4})$ & $\chi^2/3$ & $p$ & $K\bar K\pi$ & $\pi\pi\eta$ & $\pi\pi\eta'$ & $K\bar K\eta'$\\
\midrule
Flavor-singlet hybrid & $7.372^{+1.126}_{-1.137}$ & $2.003/3$ & 0.572 & $\mathbf{+0.01}$ & $\mathbf{+0.54}$ & $+1.14$ & $\mathbf{-0.64}$\\
Trigluon glueball & $5.783^{+0.939}_{-0.941}$ & $6.265/3$ & 0.0994 & $+0.65$ & $+1.44$ & $+1.50$ & $-1.23$\\
Two-gluon glueball & $6.785^{+1.228}_{-1.228}$ & $14.373/3$ & 0.00244 & $-1.04$ & $+2.02$ & $+1.16$ & $+2.80$\\
Tetraquark & $3.756^{+0.626}_{-0.755}$ & $19.292/3$ & 0.000238 & $+2.96$ & $+3.12$ & $\mathbf{+0.12}$ & $-0.88$\\
\bottomrule
\end{tabular}
\end{table*}

The seven $PPP$ final states ($K\bar K\pi$, $\pi\pi\eta$, $\pi\pi\eta'$, $K\bar K\eta'$, $K\bar K\eta$, $\eta\eta\eta$, $\eta\eta\eta'$) are evaluated at the common mass $M_X=2.359$ GeV with the same $\eta$--$\eta'$ mixing convention and exact relativistic three-body phase space. 

To compare the calculated patterns with the measured $K\bar K\pi$, $\pi\pi\eta$, $\pi\pi\eta'$, and $K\bar K\eta'$ channels, each seven-channel relative-width vector is converted into a normalized distribution,
\begin{align}
q_i&=\frac{r_i}{\sum_{j\in 3P}r_j},\qquad
\mu_i=A_{3P}q_i,\notag\\
A_{3P}&={\cal B}(J/\psi\to\gamma X)f_{3P},
\qquad f_{3P}=\frac{\Gamma_{3P}}{\Gamma_{\rm total}},
\label{eq:ppp_shape}
\end{align}
where $r_i=R_i^{PPP}$. Each distribution is fitted with the single normalization $A_{3P}$, and the upper limit on $\eta\eta\eta'$ is imposed~\cite{BESIII:2010gmv,BESIII:2023wfi,BESIII:2026rzt,BESIII:2026mvn}. Thus $A_{3P}$ is the product of the radiative-production branching fraction and the inclusive $PPP$ fraction. 

Table~\ref{tab:pppfit} shows that the flavor-singlet hybrid gives the best overall description of the measured $PPP$ channels, while the trigluon glueball remains compatible with the present data. A recent current-field-duality construction followed by Fierz bosonization likewise finds trigluon-glueball $PPP$ ratios compatible with the BESIII data~\cite{Wang:2026iyq}. The two-gluon-glueball and compact-tetraquark patterns give substantially larger $\chi^2$ values, with their largest deviations occurring in $K\bar K\eta'$ and $\pi\pi\eta$, respectively. However, this one-normalization comparison alone is not sufficient to exclude either assignment. Further tests are therefore required. The two-gluon interpretation is subjected below to an independent $VV/f_0\eta'$ test, whereas the compact-tetraquark assignment is tested through the recoupling-sensitive $\DSP$ ratio.

The channel dependence of the pulls (the signed deviations $({\rm data}-{\rm theory})/\sigma$) is also informative. The hybrid gives the smallest absolute pull in $K\bar K\pi$, $\pi\pi\eta$, and $K\bar K\eta'$, whereas the tetraquark describes $\pi\pi\eta'$ particularly well, with a pull of only $0.12$. The assignment with the smallest absolute pull therefore changes from channel to channel, revealing complementary decay patterns and leaving open the possibility that more than one configuration contributes to the physical $X(2370)$. This motivates a future coherent amplitude analysis in which relative phases and Dalitz-dependent resonant and nonresonant contributions are fitted simultaneously. Such an analysis, however, requires the characteristic decay signatures of the pure configurations to be established first. Measurements of $K\bar K\eta$ and $\eta\eta\eta$, together with a stronger limit or an eventual measurement of $\eta\eta\eta'$, would further sharpen such an analysis and enable a quantitative test of configuration mixing.

The $PPP$ comparison reveals substantial tension with the two-gluon-glueball pattern, but this comparison alone is not sufficient to exclude the assignment. An independent $VV/f_0\eta'$ constraint is therefore applied to the pure-$2g$ interpretation using the Fierz analysis of Ref.~\cite{Tan:2026uue}. Normalizing its predicted widths to the observed $f_0(980)\eta'$ mode gives
\begin{equation}
\begin{aligned}
\frac{\Gamma(\omega\phi)}{\Gamma(f_0\eta')}&=149.6,&
\frac{\Gamma(\omega\omega)}{\Gamma(f_0\eta')}&=118.5,\\
\frac{\Gamma(\phi\phi)}{\Gamma(f_0\eta')}&=36.7.&&
\end{aligned}
\label{eq:2g_vvps}
\end{equation}
Using the measured ${\cal B}(J/\psi\to\gamma X\to\gamma f_0(980)\eta'\to\gamma K_S^0K_S^0\eta')=1.31\times10^{-5}$ and only ${\cal B}[f_0(980)\to K_S^0K_S^0]\leq1$ gives the corresponding lower estimates $1.96\times10^{-3}$, $1.55\times10^{-3}$, and $4.81\times10^{-4}$ for the $\omega\phi$, $\omega\omega$, and $\phi\phi$ product branching fractions. An earlier BESIII partial-wave analysis of $J/\psi\to\gamma\phi\phi$ found only a $1.1\sigma$ significance for an added $0^{-+}$ $X(2370)$ component, suggesting substantial tension with the predicted $\phi\phi$ rate~\cite{BESIII:2016qzq}. Taken together, the unfavorable $PPP$ fit and the tension with existing $\phi\phi$ data argue against retaining the pure-$2g$ interpretation as an independent candidate.

The pure-state comparison is therefore narrowed to the compact tetraquark, flavor-singlet hybrid, and trigluon glueball. The hybrid and trigluon-glueball assignments are both compatible with the present $PPP$ data, and their predicted relative widths in the measured channels differ only by factors of order unity. A more pronounced separation can be obtained from decay-width ratios involving channels in which the leading amplitude is suppressed by color--spin--flavor recouplings.

A direct test of the compact-tetraquark assignment is provided by flavor and decay-topology recoupling in the scalar--pseudoscalar channels. The scalar--pseudoscalar channels $a_0(1450)\pi$ and $K_0^*(1430)\bar K$ are the nonstrange and strange members of the same decay class, so their common production and overall normalization cancel. The $a_0(1450)\pi$ amplitude is sensitive to coherent cancellations among the color--spin--flavor components of the compact tetraquark, while the decay remains allowed for the gluonic and hybrid configurations. We define
\begin{equation}
\DSP=\frac{\Gamma[X\to a_0(1450)\pi]}
{\Gamma[X\to K_0^*(1430)\bar K+\cc]}.
\label{eq:dsp}
\end{equation}
For the trigluon glueball, a scalar vacuum insertion selects the scalar--scalar--pseudoscalar part of the complete color--Dirac Fierz decomposition. Summing the charge states gives
\begin{equation}
\begin{aligned}
\DSP^{3g}&=\frac{3p_a}{4p_K}
\left(\frac{2}{1+r_s}\right)^2
\left(\frac{h_a}{h_K}\right)^2
\left(\frac{\mu_\pi}{\mu_K}\right)^2,\\
r_s&=\frac{\langle\bar ss\rangle}{\langle\bar qq\rangle}.
\end{aligned}
\label{eq:dsp3g}
\end{equation}
where $h_S$ and $\mu_P$ denote the scalar- and pseudoscalar-current residues, $p_a$ and $p_K$ are the decay momenta in the $a_0\pi$ and $K_0^*K$ channels, and $r_s$ is the strange-to-light quark condensate ratio. For the trigluon glueball, normalized finite-width folding gives $\DSP^{3g}=0.994$ at the central point, with a broad input scan yielding $0.447$--$1.836$. The flavor-singlet hybrid gives $\DSP^H\simeq0.85$--$0.87$ from the leading flavor amplitude $\mathcal A_H(SP)\propto\mathrm{Tr}(SP)$.

For the compact tetraquark, the same fall-apart operator and annihilation interaction are used in the seven-dimensional $I=0$, $J^{PC}=0^{-+}$ $P$-wave basis. Both the AL1 and AP1 parametrizations give $D_{SP}^{4q}=\mathcal O(10^{-3})$. Under the matched two-channel Chew--Mandelstam line-shape treatment, the corresponding values are
\begin{align}
\DSP^{3g}&\simeq0.363,\notag\\
\DSP^{4q}&\simeq6.60\times10^{-3}\ ({\rm AL1}),
\qquad6.48\times10^{-3}\ ({\rm AP1}),
\label{eq:dspstress}
\end{align}
The minimum separation factors are $54.98$ and $55.97$, respectively. Thus, $\DSP\ll1$ selects the compact-tetraquark region, whereas an order-one value leaves the trigluon-glueball and hybrid alternatives. An order-one $\DSP$, however, does not distinguish the hybrid from the trigluon glueball. A complementary test is provided by their different spin-recoupling structures through the ratio

\begin{equation}
\RphiBV=\frac{\Gamma[X\to\phi\phi]}
{\Gamma[X\to b_1(1235)\rho]}.
\label{eq:rphi}
\end{equation}
For the lowest $0^{-+}$ hybrid, the leading nonrelativistic $X\to\phi\phi$ amplitude vanishes because the relevant non-flip spin-triplet recoupling coefficient is zero, whereas $X\to b_1(1235)\rho$ remains allowed at leading order~\cite{Page:1998gz,Farina:2023oqk}. By contrast, the tensor--tensor component of the trigluon-glueball current generates the $\phi\phi$ channel at leading order. The resulting contrast makes
$\RphiBV$ directly separates the two assignments. 

For the trigluon glueball, the tensor--tensor part of the Fierz-reduced current gives
\begin{equation}
\RphiBV^{3g}=K_{\phi/BV}r_s^2
\left[\frac{(f_\phi^T)^2}{f_{b_1}^Tf_\rho^T}\right]^2,
\qquad K_{\phi/BV}=0.09244.
\label{eq:rphi3g}
\end{equation}
where $f_\phi^T$, $f_{b_1}^T$, and $f_\rho^T$ are the tensor decay constants of the respective mesons.
With all tensor-current constants evolved to a common scale~\cite{Ball:1998sk,Jansen:2009yh,Braun:2016wnx}, the usual input range gives $\RphiBV^{3g}=0.08$--$0.31$, while the conservative factorized lower edge remains $0.039$.

For the hybrid, the leading recoupling coefficient is
\begin{equation}
\left\{\begin{matrix}
\tfrac12&\tfrac12&1\\[1mm]
\tfrac12&\tfrac12&1\\[1mm]
1&1&1
\end{matrix}\right\}=0,
\label{eq:9j}
\end{equation}
while $b_1\rho$ is allowed at leading order~\cite{Page:1998gz,Farina:2023oqk}. Complete Dirac spinors first lift this zero through the double-lower-component, or $\mathcal O(v^2)$, term in the nonrelativistic expansion. Even after this correction, $\RphiBV^H$ remains below $0.006$ over the enlarged sensitivity range. The full six-dimensional overlap gives $1.48\times10^{-3}$ in the narrow-width limit and approximately $2.0\times10^{-3}$ after spectral folding.

The conservative bounds, $\RphiBV^{3g}\geq0.039$ and $\RphiBV^{H}<0.006$, remain separated by at least a factor of $6.5$. This separation is robust against a smooth common recoil form factor because the two recoil momenta differ by only $17$ MeV at $M_X=2.359$ GeV, so form-factor effects largely cancel in $\RphiBV$. Only a strong channel-dependent correction $R_{\rm true}=R_{\rm fact}|\zeta_{\rm NF}|^2$ where $|\zeta_{\rm NF}|<0.392$ would move the trigluon-glueball edge into the hybrid range.

Together the two ratios place the three unmixed assignments in separated regions of the $(\DSP,\RphiBV)$ plane. We retain deliberately wider intervals between the calculated regions and adopt the following conservative separation criteria:
\begin{equation}
\begin{array}{rcl}
\DSP<0.02 &\Rightarrow& \text{compact-}4q\text{-like},\\
\DSP>0.20,\ \RphiBV<0.01 &\Rightarrow& \text{hybrid-like},\\
\DSP>0.20,\ \RphiBV>0.03 &\Rightarrow& \text{trigluon glueball-like}.
\end{array}
\label{eq:classifier}
\end{equation}
Both ratios are experimentally accessible in radiative $J/\psi$ decays, where the common $J/\psi\to\gamma X$ production factor cancels. The intermediate ranges $0.02<\DSP<0.20$ and $0.01<\RphiBV<0.03$ are separation bands rather than statistical confidence intervals. A measured value in either range would indicate that the pure-state alternatives are insufficient and call for an amplitude-level treatment of configuration mixing, final-state interactions, or additional decay mechanisms.

\begin{figure}[t]
\centering
\includegraphics[width=\columnwidth]{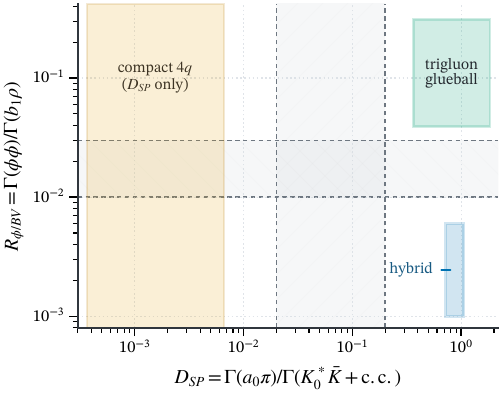}
\caption{Calculated regions and intermediate separation bands in the $(\DSP,\RphiBV)$ plane. The compact-$4q$ region is determined by $\DSP$ alone, and no prediction for its position along $\RphiBV$ is required. Hatched bands denote $0.02<\DSP<0.20$ and $0.01<\RphiBV<0.03$ and are not confidence intervals.}
\label{fig:classifier}
\end{figure}

The mass-dependent applicability of this two-ratio test can be assessed by considering the neighboring pseudoscalar candidates $X(1835)$ and $X(2120)$. Their masses admit different plausible dynamics: QCD sum rules place a $0^{-+}$ $N\bar N$ baryonium at $1.81\pm0.09$~GeV and a $0^{-+}$ trigluon glueball at $2.01\pm0.14$~GeV, close to the $X(1835)$ and $X(2120)$, respectively~\cite{Wan:2021vny,Zhang:2022obn}. Broader trigluon estimates and phenomenological arguments also motivate configuration mixing~\cite{Hao:2005hu,Ben:2026afy}. At the $X(2120)$ mass the channels entering both ratios remain open, but finite-width effects substantially erode the $\DSP$ separation and the $46$~MeV recoil-momentum mismatch weakens the cancellation in $\RphiBV$. At the $X(1835)$ mass, $K_0^*(1430)\bar K$, $\phi\phi$, and $b_1\rho$ are closed, so neither axis can be used. State-specific tests should instead combine the accessible $PPP$ amplitudes with comparative radiative-$J/\psi$ and two-photon production. These threshold and recoil limitations prevent the two-ratio strategy from being transferred unchanged to lower masses and make $X(2370)$, for which all four channels are open, particularly well suited to this test.

In conclusion, the approximately flavor-singlet assignment suggested by the $K^*\bar K+\cc$ suppression is adopted here, although this decay pattern is not unique~\cite{Ben:2026afy}; additional flavor-symmetry tests would provide a valuable cross-check. Within this framework, an ordinary light-quark-dominated $q\bar q$ pseudoscalar, predominantly $\Lambda\bar\Lambda$ baryonium, the compact-hexaquark assignment considered here, and the pure-$2g$ interpretation are disfavored as descriptions of the dominant component of $X(2370)$. Under the stated error treatment, comparison of the calculated $PPP$ decay patterns with current data favors a flavor-singlet hybrid, while a trigluon glueball remains compatible. The channel-dependent preferences leave open the possibility that more than one configuration contributes to the physical state. More decisive discrimination is provided by the complementary decay-width ratios $D_{SP}$ and $R_{\phi/BV}$, whose predictions under the compact-tetraquark, hybrid, and trigluon-glueball assignments occupy well-separated regions. Measurements of these ratios can therefore distinguish among the surviving pure-state assignments, while intermediate values would indicate the need to consider configuration mixing, final-state interactions, or additional decay mechanisms.

This work was supported by the National Key Research and Development
Program of China under Contract No. 2025YFA1613900 and by the National Natural Science Foundation of China under Grants No. 12475087, 12235008, 12575106, 12147214, and 12547114, and by the Specific Fund of Fundamental Scientific Research Operating Expenses for Undergraduate Universities in Liaoning Province under Grant No. LJ212410165019.

\onecolumngrid
\newpage
\mbox{}
\setcounter{page}{1}\renewcommand{\thepage}{S\arabic{page}}
\setcounter{equation}{0}\renewcommand{\theequation}{S\arabic{equation}}
\setcounter{table}{0}\renewcommand{\thetable}{S\arabic{table}}
\setcounter{figure}{0}\renewcommand{\thefigure}{S\arabic{figure}}
\setcounter{section}{0}\renewcommand{\thesection}{S\arabic{section}}
\begin{center}
  {\large\bfseries Supplemental Material}
\end{center}

This Supplemental Material provides the definitions, numerical inputs, consistency checks, and interpretation limits used in the Letter. The common parent parameters are
\begin{equation}
 M_X=2.359~\mathrm{GeV},
 \qquad
 \Gamma_X=0.170~\mathrm{GeV}.
 \label{eq:S_common}
\end{equation}
Sensitivity intervals quoted below are obtained by varying the parameters of the stated assignments and are not statistical confidence intervals.

\section{Hexaquark consistency tests}

Two distinct $q^3\bar q^3$ assignments are tested. For $\Lambda\bar\Lambda$ baryonium, consistency requires a state near the measured mass with a sizable projection onto the physical $\Lambda\bar\Lambda$ channel and an interaction that generates the measured complex pole. For the compact hexaquark, a single eigenstate of a Pauli-complete constituent Hamiltonian must reproduce the mass, flavor content, and measured three-pseudoscalar pattern simultaneously. These consistency tests apply only to the specific assignments considered and do not exclude an arbitrary hexaquark Fock component.

\subsection{$\Lambda\bar\Lambda$ baryonium}

\subsubsection{SU(3)-broken chromomagnetic open-door calculation}

The open-door calculation is performed for the hidden-strangeness system $qqs\bar q\bar q\bar s$ with $I=0$, total spin $J=0$, and total color singlet. Because the two light quarks are coupled to $I=0$, their color--spin wave function is symmetric; the same Pauli condition is imposed on the two light antiquarks. Direct construction in the color--spin Hilbert space leaves eight Pauli-allowed states, including the physical color-singlet $\Lambda\bar\Lambda$ direction and hidden-color directions. No projection onto the physical open channel is made before diagonalization.

The Hamiltonian is
\begin{equation}
 H_{\rm CM}
 =
 \sum_i m_i-
 \sum_{i<j}C_{ij}
 \bigl(\boldsymbol\lambda_i^c\!\cdot\!\boldsymbol\lambda_j^c\bigr)
 \bigl(\boldsymbol\sigma_i\!\cdot\!\boldsymbol\sigma_j\bigr),
 \label{eq:S_HCM}
\end{equation}
with antiquarks represented by the conjugate color generators. The basis construction and normalization follow the chromomagnetic baryon--antibaryon framework of Ref.~\cite{S-Abud:2010}. The SU(3)-broken input is
\begin{align}
 m_q&=351.65~\mathrm{MeV},&
 m_s&=455.21~\mathrm{MeV},\notag\\
 C_{qq}&=74.40~\mathrm{MeV},&
 C_{qs}&=58.04~\mathrm{MeV},&
 C_{ss}&=43.20~\mathrm{MeV}.
 \label{eq:S_HCM_inputs}
\end{align}
As a normalization check, the same operator calculation in the $J=1$ sector gives $M=2183.794$ MeV and $|c_{\Lambda\bar\Lambda}|=0.60048$, reproducing the reference result at approximately $2184$ MeV with amplitude $0.60$.

For a normalized eigenstate $|X_n\rangle$, the physical open-door probability is defined by
\begin{equation}
 P_{\Lambda\bar\Lambda}^{(n)}
 =
 \left|\langle\Lambda\bar\Lambda|X_n\rangle\right|^2.
 \label{eq:S_PLL_def}
\end{equation}
The complete $J=0$ spectrum is listed in Table~\ref{tab:S_LL_spectrum}. The state closest to $X(2370)$ and the state with the largest $\Lambda\bar\Lambda$ projection are different eigenstates:
\begin{align}
 M_{\rm near-X}&=2374.975~\mathrm{MeV},&
 P_{\Lambda\bar\Lambda}^{\rm near-X}&=0.03527,\notag\\
 M_{\Lambda\bar\Lambda\text{-rich}}&=2258.963~\mathrm{MeV},&
 P_{\Lambda\bar\Lambda}^{\rm rich}&=0.68812.
 \label{eq:S_LL_key}
\end{align}

\begin{table}[H]
\caption{Eigenvalues of the SU(3)-broken $J=0$ chromomagnetic Hamiltonian and their physical $\Lambda\bar\Lambda$ open-door probabilities.}
\label{tab:S_LL_spectrum}
\centering
\begin{tabular}{ccc}
\toprule
State & $M_n$ (MeV) & $P_{\Lambda\bar\Lambda}^{(n)}$\\
\midrule
1 & 1988.848 & $3.45\times10^{-6}$\\
2 & 2123.340 & 0.14899\\
3 & 2192.929 & 0.004456\\
4 & 2227.170 & $<10^{-28}$\\
5 & 2230.460 & 0.07860\\
6 & 2258.963 & 0.68812\\
7 & 2310.746 & 0.04456\\
8 & 2374.975 & 0.03527\\
\bottomrule
\end{tabular}
\end{table}

The stability of the near-$X$ branch is tested by varying $C_{qq}$, $C_{qs}$, and $C_{ss}$ independently by $\pm10\%$ in 3000 random samples. At each point the state is selected by maximal eigenvector overlap with the reference solution, rather than by mass ordering. The central $95\%$ scan intervals are
\begin{equation}
 2368.79<M_{\rm tracked}<2381.51~\mathrm{MeV},
 \qquad
 0.0302<P_{\Lambda\bar\Lambda}^{\rm tracked}<0.0402.
 \label{eq:S_LL_scan}
\end{equation}
The minimum squared overlap with the reference branch is 0.9953, so the identity of the eigenstate remains stable throughout the parameter variation.

To display the consequence of the different open-door projections without claiming an absolute decay width, a common residual fall-apart coupling $g_{\rm bare}$ is assigned to the two states. At the common parent mass, the reduced widths are
\begin{equation}
 \widehat\Gamma_{\rm near-X}
 =1.0743|g_{\rm bare}|^2~\mathrm{MeV},
 \qquad
 \widehat\Gamma_{\rm rich}
 =20.9570|g_{\rm bare}|^2~\mathrm{MeV},
 \label{eq:S_LL_reduced}
\end{equation}
and hence
\begin{equation}
 \frac{\widehat\Gamma_{\rm rich}}
 {\widehat\Gamma_{\rm near-X}}=19.51.
 \label{eq:S_LL_ratio}
\end{equation}
Only the ratio is used as a structural indicator. The unknown nonlocal spatial coupling and final-state interaction prevent either coefficient in Eq.~(\ref{eq:S_LL_reduced}) from being interpreted as an absolute physical width.

\subsubsection{Independent second-sheet pole test}

The open-door projection is complemented by a single-channel $\Lambda\bar\Lambda$ quasipotential Bethe--Salpeter equation (qBSE), using the same one-boson-exchange and regulator framework as the light-baryonium calculation of Ref.~\cite{S-Wan:2021vny}. In operator notation,
\begin{equation}
 T(E)=V(E)+V(E)G(E)T(E),
 \qquad
 T(E)=\bigl[1-V(E)G(E)\bigr]^{-1}V(E).
 \label{eq:S_qBSE}
\end{equation}
The interaction consists of the one-boson-exchange kernel and a Gaussian short-range optical term,
\begin{equation}
 V=V_{\rm OBE}+(U_{\rm opt}+iW_{\rm opt})V_{\rm opt},
 \label{eq:S_optical}
\end{equation}
where $W_{\rm opt}<0$ is absorptive in the convention used here. The pole to be reproduced is
\begin{equation}
 E_p=M_X-\frac{i}{2}\Gamma_X
 =2.359-i\,0.085~\mathrm{GeV}.
 \label{eq:S_target_pole}
\end{equation}
For the spectator normalization used in the numerical equation, the on-shell continuation is implemented through
\begin{equation}
 G_{\rm II}(E)-G_{\rm I}(E)
 =-\frac{ik(E)}{16\pi^2E}|k\rangle\langle k|,
 \label{eq:S_sheet_jump}
\end{equation}
and a pole satisfies
\begin{equation}
 \det\!\left[1-V(E_p)G_{\rm II}(E_p)\right]=0.
 \label{eq:S_pole_condition}
\end{equation}

On the explicitly continued sheet, the minimal-norm solution that reproduces the pole in Eq.~(\ref{eq:S_target_pole}) has $W_{\rm opt}=+0.806$ GeV and is therefore emissive rather than absorptive. The pole residue gives an elastic-equivalent width
\begin{equation}
 \Gamma_{\Lambda\bar\Lambda}^{\rm diag}
 \simeq183~\mathrm{MeV},
 \label{eq:S_LL_diag_width}
\end{equation}
which is already slightly larger than the nominal total width $\Gamma_X=170$ MeV. Equation~(\ref{eq:S_LL_diag_width}) is not quoted as a decay prediction; it tests the consistency of the imposed pole condition. Together, the small open-door projection of the near-mass eigenstate and the absence of a physically acceptable second-sheet pole disfavor the specific single-channel $\Lambda\bar\Lambda$ baryonium assignment. The conclusion does not apply to a coupled-channel baryon--antibaryon model or to a subleading baryonic component of the physical state.

\subsection{Compact hexaquark}

\subsubsection{Pauli-complete Hamiltonian and spatial basis}

The compact calculation uses a $q^3\bar q^3$ basis with
\begin{equation}
 J=0,
 \qquad I=0,
 \qquad Y=0,
 \qquad \text{total color }\mathbf 1_c.
 \label{eq:S_6q_quantum}
\end{equation}
Pauli antisymmetry is imposed separately in the three-quark and three-antiquark clusters. The resulting internal Hilbert space contains 32 Pauli-allowed states. Its total-flavor-singlet subspace is eight-dimensional: two baryon-color $\mathbf1_c\otimes\mathbf1_c$ directions and six hidden-color compact directions. The full 32-dimensional space, rather than only this singlet subspace, is diagonalized so that SU(3) breaking and mixing among allowed total-flavor representations are retained.

Two standard constituent-quark interactions, AL1 and AP1 \cite{S-Semay:1994ht}, are used as independent parametrizations. Their pair potential is
\begin{align}
 V_{ij}(r)=&-\frac{3}{16}
 \bigl(\boldsymbol\lambda_i^c\!\cdot\!\boldsymbol\lambda_j^c\bigr)
 \left[
 \lambda r^p-\frac{\kappa}{r}-\Lambda
 \right.\notag\\
 &\left.
 +\frac{2\pi\kappa'}{3m_im_j}
 \frac{e^{-r^2/r_{0,ij}^2}}
 {\pi^{3/2}r_{0,ij}^3}
 \bigl(\boldsymbol\sigma_i\!\cdot\!\boldsymbol\sigma_j\bigr)
 \right],
 \label{eq:S_AL1AP1}
\end{align}
with
\begin{equation}
 r_{0,ij}=A
 \left(\frac{2m_im_j}{m_i+m_j}\right)^{-B}.
 \label{eq:S_r0}
\end{equation}
The numerical inputs are collected in Table~\ref{tab:S_AL1AP1}.

\begin{table}[H]
\caption{AL1/AP1 inputs used in the compact-hexaquark calculation. Masses and $\Lambda$ are in GeV; the remaining dimensions follow from Eq.~(\ref{eq:S_AL1AP1}).}
\label{tab:S_AL1AP1}
\centering
\begin{tabular}{lcccccccc}
\toprule
Model & $p$ & $\kappa$ & $\kappa'$ & $\lambda$ & $\Lambda$ & $A$ & $B$ & $(m_n,m_s)$\\
\midrule
AL1 & 1 & 0.5069 & 1.8609 & 0.1653 & 0.8321 & 1.6553 & 0.2204 & $(0.315,0.577)$\\
AP1 & $2/3$ & 0.4242 & 1.8025 & 0.3898 & 1.1313 & 1.5296 & 0.3263 & $(0.277,0.553)$\\
\bottomrule
\end{tabular}
\end{table}

The five internal Jacobi coordinates are represented by a two-scale Gaussian expansion,
\begin{equation}
 \Phi_{ab}(\boldsymbol\xi_1,\ldots,\boldsymbol\xi_5)
 ={\cal N}_{ab}
 \exp\!\left[-\frac12
 \left(a\sum_{k=1}^{4}\xi_k^2+b\xi_5^2\right)\right],
 \label{eq:S_6q_gaussian}
\end{equation}
where $a$ controls the four intracluster coordinates and $b$ the relative coordinate between the clusters. Seven geometrically spaced values of each width between $0.025$ and $1.2~\mathrm{GeV}^2$ are used in the central calculation. The variational coefficients follow from
\begin{equation}
 H\bm c=E\,N\bm c,
 \label{eq:S_6q_GEVP}
\end{equation}
after overlap eigenvalues below $10^{-10}$ of the largest eigenvalue are removed. Increasing the spatial basis from $6\times6$ to $7\times7$ changes the tracked compact level by $-1.23$ MeV for AL1 and $-0.46$ MeV for AP1, while its internal probabilities change only at the $10^{-3}$ level.

Total-flavor probabilities are obtained from spectral projectors of the SU(3) quadratic Casimir,
\begin{equation}
 P_R^{(n)}
 =\langle X_n|{\cal P}_R|X_n\rangle,
 \qquad
 {\cal P}_R
 =\sum_{\alpha\in R}|\alpha\rangle\langle\alpha|,
 \label{eq:S_6q_flavor_projector}
\end{equation}
where $R=\mathbf1,\mathbf8,\mathbf{27},\mathbf{64}$ in the present sector. The baryon-color probability $P_{B\bar B}$ is defined analogously by projection onto the $\mathbf1_c\otimes\mathbf1_c$ cluster-color subspace.

Before flavor-changing annihilation is included, the state tracked near $X(2370)$ has the composition shown in Table~\ref{tab:S_6q_H0}. Its mass and four measured $PPP$ ratios appear acceptable, but its total-flavor wave function is predominantly octet rather than singlet.

\begin{table}[H]
\caption{Uncalibrated near-$X$ compact-hexaquark branch. The last column gives the four-channel one-normalization comparison defined below.}
\label{tab:S_6q_H0}
\centering
\begin{tabular}{lcccccc}
\toprule
Model & $M$ (MeV) & $P_1$ & $P_8$ & $P_{27}$ & $P_{64}$ & $\chi^2_{4PPP}$\\
\midrule
AL1 & 2378.423 & 0.37478 & 0.59840 & 0.02590 & 0.00091 & 2.83\\
AP1 & 2394.425 & 0.37469 & 0.59773 & 0.02628 & 0.00130 & 3.14\\
\bottomrule
\end{tabular}
\end{table}

\subsubsection{Independent $\eta$--$\eta'$ calibration of flavor-changing annihilation}

The flavor-changing interaction is fixed without using the mass or decay data of $X(2370)$. For each AL1/AP1 Hamiltonian, the bare $n\bar n$ and $s\bar s$ pseudoscalar ground states are first solved. In the $(\eta_q,\eta_s)$ basis, with $\eta_q=(u\bar u+d\bar d)/\sqrt2$, the target mass matrix is
\begin{equation}
 H_{\eta\eta'}^{\rm phys}
 =R(\phi)
 \begin{pmatrix}m_\eta&0\\0&m_{\eta'}\end{pmatrix}
 R^T(\phi),
 \qquad \phi=39.3^\circ,
 \label{eq:S_eta_target}
\end{equation}
where $m_\eta=547.862$ MeV and $m_{\eta'}=957.780$ MeV. The required annihilation matrix is
\begin{equation}
 A_{\eta\eta'}
 =H_{\eta\eta'}^{\rm phys}
 -\operatorname{diag}(M_{\eta_q}^{(0)},M_{\eta_s}^{(0)}).
 \label{eq:S_eta_ann_matrix}
\end{equation}

For an annihilation range $r_A$, let $F_{nn}$, $F_{ns}$, and $F_{ss}$ be the Gaussian contact integrals between the corresponding meson wave functions. The three flavor couplings are determined algebraically by
\begin{equation}
 g_N=\frac{(A_{\eta\eta'})_{qq}}{2F_{nn}},
 \qquad
 g_{NS}=\frac{(A_{\eta\eta'})_{qs}}{\sqrt2F_{ns}},
 \qquad
 g_S=\frac{(A_{\eta\eta'})_{ss}}{F_{ss}}.
 \label{eq:S_eta_couplings}
\end{equation}
The resulting pair interaction is embedded in the hexaquark Hamiltonian as
\begin{equation}
 V_A(r_A)=
 \sum_{i=1}^{3}\sum_{j=4}^{6}
 f_A(r_{ij};r_A)
 {\cal P}^{(ij)}_{c=1}
 {\cal P}^{(ij)}_{S=0}
 {\cal G}^{(ij)}_F,
 \label{eq:S_6q_annihilation}
\end{equation}
so that annihilation acts only on color-singlet, spin-singlet $q\bar q$ pairs. At every $r_A$ the meson calibration is repeated before the full hexaquark Hamiltonian is rediagonalized. The range is not fitted to $X(2370)$: the scan $r_A=0.30$--$0.49$ fm surrounds the AL1/AP1 hyperfine-smearing scales, with central values $0.398$ fm (AL1) and $0.418$ fm (AP1).

The calibrated Hamiltonian is diagonalized in an increasing set of unperturbed eigenstates. Raising the retained-energy cutoff from $2.7$ to $4.5$ GeV increases the basis to 234 states for AL1 and 257 for AP1. Above a $4.0$ GeV cutoff, the tracked-branch masses change by less than $0.4$ MeV (AL1) and $0.8$ MeV (AP1). The original near-$X$ branch is followed by maximum overlap with its uncalibrated eigenvector; a separate state is selected solely by proximity to $M_X$ after rediagonalization. This distinction prevents an avoided crossing or eigenvalue reordering from being mistaken for survival of the original state.

The converged central results are summarized in Table~\ref{tab:S_6q_calibrated}. The branch continuously connected by overlap to the uncalibrated near-$X$ solution moves to $2.61$--$2.63$ GeV. Its overlap with the original state is only $0.27$--$0.29$, reflecting strong rearrangement. The newly selected near-mass state has an overlap below $1.3\%$ with the original branch, only $23$--$24\%$ total-flavor-singlet probability, and a large baryon-color projection.

\begin{table}[H]
\caption{Central calibrated compact-hexaquark results at the converged $4.5$ GeV unperturbed-basis cutoff. ``Tracked'' denotes maximum overlap with the uncalibrated near-$X$ branch; ``near'' denotes the state closest to $M_X$ after rediagonalization.}
\label{tab:S_6q_calibrated}
\centering
\begin{tabular}{llccccc}
\toprule
Model & State & $M$ (MeV) & $|\langle X^{(0)}|X\rangle|^2$ & $P_1$ & $P_{B\bar B}$ & $\chi^2_{4PPP}$\\
\midrule
AL1 & tracked & 2632.793 & 0.27463 & 0.28041 & 0.53068 & 6.95\\
AL1 & near & 2340.559 & 0.01274 & 0.23431 & 0.55369 & 29.28\\
AP1 & tracked & 2609.556 & 0.28764 & 0.32040 & 0.67317 & 5.20\\
AP1 & near & 2356.572 & 0.00493 & 0.24224 & 0.65262 & 25.55\\
\bottomrule
\end{tabular}
\end{table}

Neither the tracked branch nor the state remaining near $X(2370)$ is singlet dominated after calibration. The original near-$X$ solution is instead displaced and strongly rearranged, while the replacement near-mass state is not a continuation of it.

\subsubsection{$PPP$ doorway projection and decay self-consistency}
\label{sec:S_6q_PPP}

The decay calculation uses the actual SU(3)-broken 32-dimensional eigenvector, not a pure-flavor tensor. For every charge channel $f$, a color-singlet, spin-singlet three-meson doorway $|d_f\rangle$ is constructed by pairing the three quarks with the three antiquarks and attaching the same pseudoscalar-current residues and $\eta$--$\eta'$ convention used for the other $PPP$ calculations. If the normalized eigenvector is written as $C_{\mu\alpha}$ in an orthonormal spatial basis $\mu$ and internal basis $\alpha$, its inclusive doorway strength is
\begin{equation}
 S_f=
 \sum_\mu
 \left|\sum_\alpha C_{\mu\alpha}d_{f,\alpha}^*\right|^2.
 \label{eq:S_6q_doorway}
\end{equation}
This expression preserves coherent interference among internal color--spin--flavor components at fixed spatial index while summing over the unresolved spatial final state. The relative widths are
\begin{equation}
 R_f^{PPP}=\rho_f\frac{S_f}{S_{K\bar K\pi}},
 \label{eq:S_6q_RPPP}
\end{equation}
where $\rho_f$ is the exact relativistic three-body phase-space ratio, including charge multiplicities and identical-particle factors, at $M_X=2.359$ GeV. Because a nonlocal three-meson continuum kernel is not included, Eq.~(\ref{eq:S_6q_RPPP}) gives a doorway-level estimate of the relative decay widths rather than an exclusive absolute-width calculation.

The four measured product branching fractions are represented by
\begin{equation}
 \bm d=(3.25,3.20,1.94,0.39)\times10^{-4}
 \label{eq:S_6q_data}
\end{equation}
for $(K\bar K\pi,\pi\pi\eta,\pi\pi\eta',K\bar K\eta')$. For a fixed theoretical shape, only a common normalization $\mu$ is fitted:
\begin{equation}
 \chi^2_{4PPP}(\mu)=
 \sum_i
 \frac{\left(\mu R_i^{PPP}-d_i\right)^2}
 {\sigma_i^2(\mu)},
 \label{eq:S_6q_chi2}
\end{equation}
where the upper or lower combined experimental uncertainty is selected according to the side of the central value on which the prediction lies. The same data and uncertainty prescription are used at every Hamiltonian point; no channel-dependent normalization is introduced.

Before annihilation is calibrated, the near-$X$ branches give $\chi^2_{4PPP}=2.83$ (AL1) and $3.14$ (AP1). After calibration, the near-mass states give the ratios in Table~\ref{tab:S_6q_PPP_final}. The large $\pi\pi\eta$ and $\pi\pi\eta'$ enhancements drive $\chi^2_{4PPP}=29.28$ and $25.55$, while the predicted $\eta\eta\eta'$ ratios also lie well above the experimental upper limit $R_{\eta\eta\eta'}^{PPP}<0.0283$.

\begin{table}[H]
\caption{Calibrated near-mass compact-hexaquark $PPP$ doorway ratios, normalized to $R_{K\bar K\pi}^{PPP}=1$.}
\label{tab:S_6q_PPP_final}
\centering
\begin{tabular}{lccccc}
\toprule
Model & $R_{\pi\pi\eta}$ & $R_{\pi\pi\eta'}$ & $R_{K\bar K\eta'}$ & $R_{\eta\eta\eta}$ & $R_{\eta\eta\eta'}$\\
\midrule
AL1 & 4.27997 & 6.99330 & 0.03480 & 0.50219 & 0.16310\\
AP1 & 7.14230 & 5.84268 & 0.02766 & 0.61594 & 0.18297\\
\bottomrule
\end{tabular}
\end{table}

Over $r_A=0.30$--$0.49$ fm, the near-mass state remains incompatible with the measured $PPP$ pattern: $\chi^2_{4PPP}=27.7$--$30.1$ for AL1 and $24.7$--$27.2$ for AP1. The failure is therefore not tied to one annihilation range. The compact-hexaquark assignment loses simultaneous consistency: the uncalibrated near-$X$ branch has an unsuitable flavor composition; the independently calibrated interaction displaces and rearranges that branch; and the state that remains near the measured mass fails the observed $PPP$ pattern. This conclusion applies to the specific Pauli-complete AL1/AP1 assignment and not to an arbitrary hexaquark admixture.

\section{Unified three-pseudoscalar decay patterns}
\label{sec:S_ppp}

\subsection{Common conventions and three-body phase space}

For a charge-specific final state $f=P_1P_2P_3$, the Lorentz-invariant phase space is
\begin{equation}
 d\Phi_3=(2\pi)^4\delta^{(4)}\!\left(P-\sum_{i=1}^3p_i\right)
 \prod_{i=1}^3\frac{d^3\bm p_i}{(2\pi)^3,2E_i},
 \label{eq:S_ppp_phase_space}
\end{equation}
and the width is
\begin{equation}
 \Gamma_f=\frac{1}{2M_XS_f}\int d\Phi_3\,
 |{\cal M}_f(s_{12},s_{23})|^2.
 \label{eq:S_ppp_width}
\end{equation}
Here $S_f$ is the product of factorials for identical particles in that charge mode. After the trivial angular integrations, Eq.~(\ref{eq:S_ppp_width}) can be evaluated as
\begin{equation}
 \Gamma_f=\frac{1}{256\pi^3M_X^3S_f}
 \int_{s_{12}^{\rm min}}^{s_{12}^{\rm max}}ds_{12}
 \int_{s_{23}^-}^{s_{23}^+}ds_{23}\,
 |{\cal M}_f(s_{12},s_{23})|^2,
 \label{eq:S_ppp_dalitz}
\end{equation}
where
\begin{equation}
 s_{12}^{\rm min}=(m_1+m_2)^2,
 \qquad
 s_{12}^{\rm max}=(M_X-m_3)^2,
\end{equation}
and
\begin{align}
 s_{23}^{\pm}={}&m_2^2+m_3^2+
 \frac{1}{2s_{12}}
 \Bigl[(M_X^2-m_3^2-s_{12})(s_{12}+m_2^2-m_1^2)
 \notag\\
 &\hspace{20mm}\pm
 \lambda^{1/2}(s_{12},m_1^2,m_2^2)
 \lambda^{1/2}(M_X^2,s_{12},m_3^2)\Bigr].
 \label{eq:S_ppp_boundaries}
\end{align}
The members of a quoted channel family are summed only after the charge-specific phase-space integral and identical-particle factor have been applied. All results use $M_X=2.359$ GeV and are normalized after integration according to
\begin{equation}
 R_f^{PPP}=\frac{\Gamma_f}{\Gamma(K\bar K\pi)},
 \qquad R_{K\bar K\pi}^{PPP}=1.
 \label{eq:S_ppp_def}
\end{equation}
The common channel order is
\begin{equation}
 \bm R^{PPP}=\bigl(R_{K\bar K\pi},R_{\pi\pi\eta},R_{\pi\pi\eta'},
 R_{K\bar K\eta},R_{K\bar K\eta'},R_{\eta\eta\eta},
 R_{\eta\eta\eta'}\bigr).
 \label{eq:S_ppp_order}
\end{equation}

The quark-flavor convention is
\begin{equation}
 \begin{pmatrix}\eta\\ \eta'\end{pmatrix}
 =
 \begin{pmatrix}\cos\phi&-\sin\phi\\ \sin\phi&\cos\phi\end{pmatrix}
 \begin{pmatrix}\eta_q\\ \eta_s\end{pmatrix},
 \qquad
 \eta_q=\frac{u\bar u+d\bar d}{\sqrt2},
 \qquad \phi=39.3^\circ,
 \label{eq:S_ppp_eta_mix}
\end{equation}
For the determinant construction, we use the pseudoscalar-field
renormalization factors of the extended Linear Sigma Model
(eLSM)~\cite{S-Eshraim:2012kr,S-Giacosa:2024sgv},
\begin{equation}
 Z_\pi=1.709,
 \qquad Z_K=1.604,
 \qquad Z_{\eta_s}=1.539.
 \label{eq:S_ppp_Z}
\end{equation}
The calculated relative-width patterns in this section do not include an isobar decomposition, finite-width convolution of an intermediate resonance, or coherent interference between Dalitz amplitudes. Their scope is therefore the integrated flavor pattern in the stated decay operator.

\subsection{Decay operators and calculated $PPP$ patterns}

For the two-gluon glueball, we adopt the anomalous determinant interaction first
applied to pseudoscalar-glueball decays into scalar and pseudoscalar mesons in
Ref.~\cite{S-Eshraim:2012kr}:
\begin{equation}
 {\cal L}_{GPPP}=ic_G\widetilde G
 \left(\det\Phi-\det\Phi^\dagger\right),
 \qquad \Phi=S+iP,
 \label{eq:S_ppp_2g}
\end{equation}
Determinant-like anomalous interactions involving mesons with nonzero spin and
glueballs were subsequently developed in an instanton-motivated framework in
Ref.~\cite{S-Giacosa:2023lmi}. The $PPP$ term of the interaction above is
proportional to $\widetilde G\det P$. In the quark-flavor basis,
\begin{equation}
 P=\begin{pmatrix}
 (\pi^0+\eta_q)/\sqrt2&\pi^+&K^+\\
 \pi^-&(-\pi^0+\eta_q)/\sqrt2&K^0\\
 K^-&\bar K^0&\eta_s
 \end{pmatrix},
 \label{eq:S_ppp_Pmatrix}
\end{equation}
followed by the transformation in Eq.~(\ref{eq:S_ppp_eta_mix}) and the field renormalizations in Eq.~(\ref{eq:S_ppp_Z}). Expanding the determinant fixes the relative amplitude of every charge mode; the common constant $c_G$ cancels from Eq.~(\ref{eq:S_ppp_def}).

For the trigluon-glueball, the relevant part of the Fierz-reduced current is
\begin{equation}
 -\widetilde J_{3g}\big|_{PPP}=\frac{1}{3}{\rm Tr}(P^3).
 \label{eq:S_ppp_3g}
\end{equation}
The trace is expanded before charge summation, and the resulting flavor coefficients multiply the exact phase-space integrals in Eq.~(\ref{eq:S_ppp_width}). The overall current normalization cancels in the ratios. The flavor-singlet hybrid pattern is obtained from its single-trace flavor amplitude with flavor-blind pair creation. Only the orderings allowed by the hybrid decay topology are retained; the common pair-creation strength again cancels. This flavor-level calculation is separate from the spin-dependent hybrid calculation used below for $R_{\phi/BV}$.

The compact-$4q$ pattern in this section is obtained from an interpolating-current calculation for the physical compact $P$-wave tetraquark assignment considered in the Letter. It provides the inclusive seven-channel $PPP$ pattern, whereas the seven-dimensional eigenstate calculation below resolves the same assignment into the specific scalar--pseudoscalar channels entering $D_{SP}$. For the compact hexaquark, the complete 32-dimensional eigenvector is projected through the color--spin--flavor $q^3\bar q^3\to PPP$ doorway described in the $PPP$ doorway-projection subsection above. No neighboring-channel interpolation is used when a projection has not been calculated.

\begin{table*}[t]
\caption{The calculated $PPP$ patterns and their role in the Letter. ``Seven-channel'' means complete only within the specified operator; it does not denote a calculation of the full coherent Dalitz amplitude.}
\label{tab:S_ppp_status}
\centering
\scriptsize
\begin{tabular}{lll}
\toprule
Configuration & Calculation & Use in the Letter\\
\midrule
Two-gluon glueball & determinant, seven channels & inclusive $PPP$ comparison\\
Trigluon-glueball & current/Fierz, seven channels & comparison and one-normalization fit\\
Flavor-singlet hybrid & single trace, seven channels & comparison and one-normalization fit\\
Compact $4q$ & interpolating current, seven channels & inclusive $PPP$ comparison\\
Compact hexaquark & full-32D state projection & mass--flavor--decay test\\
$q\bar q$ and baryonium & no common calculation & not entered in the $PPP$ fit\\
\bottomrule
\end{tabular}
\end{table*}

\subsection{Relative widths and experimental data}

The available experimental product branching fractions, in units of $10^{-4}$, are
\begin{table}[H]
\caption{Experimental inputs used in the one-normalization comparison. The statistical and systematic uncertainties have been combined in quadrature separately above and below the central value.}
\label{tab:S_ppp_data}
\centering
\begin{tabular}{lccc}
\toprule
Channel & $d_i$ & $\sigma_{+,i}$ & $\sigma_{-,i}$\\
\midrule
$K\bar K\pi$ & 3.25 & 0.772 & 0.791\\
$\pi\pi\eta$ & 3.20 & 0.906 & 1.005\\
$\pi\pi\eta'$ & 1.94 & 0.332 & 0.881\\
$K\bar K\eta'$ & 0.39 & 0.112 & 0.112\\
\bottomrule
\end{tabular}
\end{table}
The additional constraint is
\begin{equation}
 {\cal B}(J/\psi\to\gamma X)
 {\cal B}(X\to\eta\eta\eta')<9.2\times10^{-6}
 \qquad(90\%\ {\rm C.L.}).
 \label{eq:S_ppp_UL}
\end{equation}
The corresponding normalized experimental pattern is
\begin{equation}
 \bm R_{\rm exp}^{PPP}=(1,0.985,0.597,{\rm n.m.},0.120,
 {\rm n.m.},<0.0283),
 \label{eq:S_ppp_expvector}
\end{equation}
where ``n.m.'' denotes not measured. In the theoretical rows, ``n.c.'' instead denotes that no result was calculated in the stated framework; it must not be read as a zero width.

\begin{table}[H]
\caption{Relative $PPP$ widths normalized to $K\bar K\pi=1$. The compact-$4q$ row is the local-current result, whereas the compact-hexaquark rows use the calibrated full-32D near-mass states.}
\label{tab:S_ppp}
\centering
\scriptsize
\setlength{\tabcolsep}{3.5pt}
\begin{tabular}{lrrrrrrr}
\toprule
Configuration & $K\bar K\pi$ & $\pi\pi\eta$ & $\pi\pi\eta'$ & $K\bar K\eta$ & $K\bar K\eta'$ & $\eta\eta\eta$ & $\eta\eta\eta'$\\
\midrule
Experimental ratios & 1 & 0.985 & 0.597 & n.m. & 0.120 & n.m. & $<0.0283$\\
Two-gluon determinant & 1 & 0.2886 & 0.2266 & 0.1072 & 0.01887 & 0.03167 & 0.000352\\
Trigluon current & 1 & 0.640623 & 0.225725 & 0.013476 & 0.192841 & 0.010238 & 0.030330\\
Singlet hybrid & 1 & 0.8192 & 0.2883 & 0.002670 & 0.1423 & 0.000505 & 0.01810\\
$4q$ local current & 1 & 0.0704 & 2.0162 & 0.3415 & 0.5371 & 0.1374 & 0.0258\\
Compact hexaquark, AL1 & 1 & 4.2800 & 6.9933 & n.c. & 0.03480 & 0.5022 & 0.1631\\
Compact hexaquark, AP1 & 1 & 7.1423 & 5.8427 & n.c. & 0.02766 & 0.6159 & 0.1830\\
\bottomrule
\end{tabular}
\end{table}

For later use, the four complete theoretical distributions within the seven-channel $PPP$ set are
\begin{equation}
 q_i=\frac{R_i^{PPP}}{\sum_{j\in3P}R_j^{PPP}},
 \qquad \sum_iq_i=1.
 \label{eq:S_ppp_q}
\end{equation}
Their numerical values are
\begin{table}[H]
\caption{Conditional fractions $q_i$ within the seven $PPP$ channels. They are not branching fractions relative to the total $X$ width.}
\label{tab:S_ppp_q}
\centering
\scriptsize
\setlength{\tabcolsep}{4pt}
\begin{tabular}{lrrrrrrr}
\toprule
Configuration & $K\bar K\pi$ & $\pi\pi\eta$ & $\pi\pi\eta'$ & $K\bar K\eta$ & $K\bar K\eta'$ & $\eta\eta\eta$ & $\eta\eta\eta'$\\
\midrule
Two-gluon determinant & 0.5976 & 0.1725 & 0.1354 & 0.0641 & 0.0113 & 0.0189 & 0.0002\\
Trigluon current & 0.4732 & 0.3031 & 0.1068 & 0.0064 & 0.0913 & 0.0048 & 0.0144\\
Singlet hybrid & 0.4403 & 0.3607 & 0.1269 & 0.0012 & 0.0627 & 0.0002 & 0.0080\\
$4q$ local current & 0.2422 & 0.0171 & 0.4884 & 0.0827 & 0.1301 & 0.0333 & 0.0062\\
\bottomrule
\end{tabular}
\end{table}

The ratios of the trigluon-glueball to hybrid predictions in the three principal nonreference modes are
\begin{equation}
 \frac{R_{\pi\pi\eta}^{3g}}{R_{\pi\pi\eta}^{H}}=0.782,
 \qquad
 \frac{R_{\pi\pi\eta'}^{3g}}{R_{\pi\pi\eta'}^{H}}=0.783,
 \qquad
 \frac{R_{K\bar K\eta'}^{3g}}{R_{K\bar K\eta'}^{H}}=1.36.
 \label{eq:S_ppp_3gH}
\end{equation}
All are of order unity. The larger contrast in $\eta\eta\eta$ occurs between two suppressed widths,
\begin{equation}
 R_{\eta\eta\eta}^{3g}=1.02\times10^{-2},
 \qquad
 R_{\eta\eta\eta}^{H}=5.05\times10^{-4},
\end{equation}
which are more exposed to subleading operators, intermediate resonances, and experimental sensitivity. This is why no inclusive $PPP$ ratio is used as the final structure-sensitive observable.

\subsection{One-normalization fit and incoherent combinations}

For a theoretical distribution $q_i$, the predicted product branching fraction is written as
\begin{equation}
 \mu_i=A_{3P}q_i,
 \qquad
 A_{3P}={\cal B}(J/\psi\to\gamma X)f_{3P},
 \qquad
 f_{3P}=\frac{\Gamma_{3P}}{\Gamma_{\rm total}}.
 \label{eq:S_ppp_A3P}
\end{equation}
Thus the one fitted parameter is $A_{3P}$, not the absolute radiative production branching fraction. For the four channels with central values, asymmetric errors are treated with
\begin{equation}
 \chi_i^2(A_{3P})=
 \begin{cases}
 (d_i-A_{3P}q_i)^2/\sigma_{+,i}^2,&A_{3P}q_i>d_i,\\
 (d_i-A_{3P}q_i)^2/\sigma_{-,i}^2,&A_{3P}q_i\le d_i.
 \end{cases}
 \label{eq:S_ppp_split}
\end{equation}
For the upper limit in Eq.~(\ref{eq:S_ppp_UL}), the one-sided contribution is
\begin{equation}
 \chi_{\rm UL}^2=
 \frac{[\max(0,\mu_{\eta\eta\eta'}-0.092)]^2}{\sigma_{\rm UL}^2},
 \qquad
 \sigma_{\rm UL}=\frac{0.092}{1.28155}=0.0718,
 \label{eq:S_ppp_UL_penalty}
\end{equation}
in units of $10^{-4}$. A continuous one-dimensional minimization is used, and the quoted $1\sigma$ interval of $A_{3P}$ is defined by $\Delta\chi^2=1$. With four central values and one fitted normalization, the nominal number of degrees of freedom is three. The upper-limit term is inactive at all four minima. Signed pulls are $(d_i-\mu_i)/\sigma_i$ with the same side-dependent error as in Eq.~(\ref{eq:S_ppp_split}). Experimental correlations and a common theory covariance are unavailable; the resulting $p$ values therefore describe compatibility only in this one-normalization comparison with diagonal errors.

The channel-by-channel pattern is more informative than the ordering of the total $\chi^2$. The hybrid has the smallest total $\chi^2$ and the smallest absolute pull in $K\bar K\pi$, $\pi\pi\eta$, and $K\bar K\eta'$. The local-current $4q$ result has the largest total $\chi^2$, but it gives the smallest pull in $\pi\pi\eta'$, namely $0.12$. The best-described channel therefore changes with the assumed configuration, showing that different channels have complementary sensitivities to the assumed constituent structures.

For illustration, incoherent sums of the integrated decay distributions are defined by
\begin{equation}
 q_i^{\rm mix}=\sum_\alpha w_\alpha q_i^{(\alpha)},
 \qquad w_\alpha\ge0,
 \qquad \sum_\alpha w_\alpha=1.
 \label{eq:S_ppp_noncoherent}
\end{equation}
For the three shapes other than the already best-fitting hybrid, the results are
\begin{table*}[t]
\caption{Incoherent sums of integrated $PPP$ distributions. The weights are not configuration probabilities.}
\label{tab:S_ppp_mix}
\centering
\small
\begin{tabular}{lccl}
\toprule
Combination & $\chi^2_{\min}$ & $A_{3P}\ (10^{-4})$ & Shape weights\\
\midrule
$3g$ & 6.265 & 5.783 & $w_{3g}=1$\\
$4q$ local current & 19.292 & 3.756 & $w_{4q}=1$\\
$2g$ determinant & 14.373 & 6.785 & $w_{2g}=1$\\
$3g+4q$ & 6.265 & 5.783 & $(w_{3g},w_{4q})=(1,0)$\\
$3g+2g$ & 3.981 & 7.016 & $(w_{3g},w_{2g})=(0.617,0.383)$\\
$4q+2g$ & 5.415 & 7.587 & $(w_{4q},w_{2g})=(0.327,0.673)$\\
$3g+4q+2g$ & 3.871 & 7.159 & $(w_{3g},w_{4q},w_{2g})=(0.490,0.081,0.429)$\\
\bottomrule
\end{tabular}
\end{table*}
These weights mix already integrated widths and contain neither relative phases nor interference terms. They cannot be interpreted as configuration admixtures. A physical analysis requires
\begin{equation}
 {\cal M}_i(z)=\sum_\alpha c_\alpha{\cal M}_{\alpha i}(z),
 \qquad
 \Gamma_i=\frac{1}{2M_X}\int d\Phi_i\,|{\cal M}_i(z)|^2,
 \label{eq:S_ppp_coherent_mix}
\end{equation}
with the isobar content, relative phases, Dalitz dependence, and experimental covariance included. Measurements of $K\bar K\eta$ and $\eta\eta\eta$, together with a more restrictive $\eta\eta\eta'$ limit or a central value, would provide the additional independent information needed for such an analysis.

\section{The scalar--pseudoscalar ratio}
\label{sec:S_dsp}

The charge-complete scalar--pseudoscalar ratio used in the Letter is
\begin{equation}
 \DSP=\frac{\displaystyle\sum_{c\in a_0\pi}\Gamma(X\to c)}
 {\displaystyle\sum_{c\in K_0^*\bar K+\cc}\Gamma(X\to c)}.
 \label{eq:S_dsp_definition}
\end{equation}
The numerator contains $a_0^+\pi^-$, $a_0^-\pi^+$, and $a_0^0\pi^0$, whereas the denominator contains $K_0^{*+}K^-$, $K_0^{*0}\bar K^0$, $K_0^{*-}K^+$, and $\bar K_0^{*0}K^0$. Both are $0^++0^-$ channels in relative $S$ wave. Their ratio therefore compares nonstrange and strange members of the same decay class while canceling the common parent normalization. In the compact-tetraquark calculation the numerator is particularly sensitive to coherent color--spin--flavor recoupling, as shown explicitly below.

\subsection{Trigluon-glueball and hybrid amplitudes}

The local pseudoscalar trigluon current and its current--field-duality representation are
\begin{equation}
 J_{3g}=g_s^3f^{abc}\widetilde G^a_{\mu\nu}
 \widetilde G^b_{\nu\rho}\widetilde G^c_{\rho\mu},
 \qquad
 \widetilde J^a_{\mu\nu}=\frac12\epsilon_{\mu\nu\alpha\beta}
 \bar qT^a\sigma^{\alpha\beta}q.
 \label{eq:S_dsp_current}
\end{equation}
The color reduction
\begin{equation}
 f_{abc}T^a_{ij}T^b_{kl}T^c_{mn}
 =\frac{i}{4}\left(\delta_{in}\delta_{kj}\delta_{ml}
 -\delta_{il}\delta_{kn}\delta_{mj}\right)
 \label{eq:S_dsp_color}
\end{equation}
and the complete Clifford rearrangement give the scalar--scalar--pseudoscalar term
\begin{equation}
 -\widetilde J_{3g}\big|_{SSP}=-{\rm Tr}(S^2P).
 \label{eq:S_dsp_SSP}
\end{equation}
With
\begin{equation}
 \Sigma=\langle\bar qq\rangle\,{\rm diag}(1,1,r_s),
 \qquad r_s=\frac{\langle\bar ss\rangle}{\langle\bar qq\rangle},
\end{equation}
one scalar vacuum insertion yields
\begin{equation}
 -\widetilde J_{3g}\big|_{SP}^{\rm vac}
 \propto-{\rm Tr}(\{\Sigma,S\}P).
 \label{eq:S_dsp_vac}
\end{equation}
The light--light $a_0\pi$ amplitudes carry the coefficient $2\langle\bar qq\rangle$, while the light--strange $K_0^*K$ amplitudes carry $(1+r_s)\langle\bar qq\rangle$. Explicitly,
\begin{align}
 {\cal O}_{SP}^{3g}\propto{}&2\langle\bar qq\rangle
 (a_0^+\pi^-+a_0^-\pi^++a_0^0\pi^0)\notag\\
 &+(1+r_s)\langle\bar qq\rangle
 (K_0^{*+}K^-+K_0^{*0}\bar K^0+K_0^{*-}K^+
 +\bar K_0^{*0}K^0)+\cdots .
 \label{eq:S_dsp_charges}
\end{align}

Defining
\begin{equation}
 \langle0|S_a|a_0\rangle=h_a,
 \quad \langle0|S_K|K_0^*\rangle=h_K,
 \quad \langle0|P_\pi|\pi\rangle=\mu_\pi,
 \quad \langle0|P_K|K\rangle=\mu_K,
\end{equation}
and using the $S$-wave two-body width, all common current and Fierz normalizations cancel, leaving
\begin{equation}
 \boxed{
 \DSP^{3g}=\frac{3p_a}{4p_K}
 \left|\frac{2}{1+r_s}\right|^2
 \left|\frac{h_a}{h_K}\right|^2
 \left|\frac{\mu_\pi}{\mu_K}\right|^2.}
 \label{eq:S_dsp_3g}
\end{equation}
At the nominal masses,
\begin{equation}
 p_{a_0\pi}=0.73183~{\rm GeV},
 \qquad p_{K_0^*K}=0.62945~{\rm GeV},
 \qquad \frac{3p_{a_0\pi}}{4p_{K_0^*K}}=0.87199.
\end{equation}
Using $\mu_\pi/\mu_K=0.90764$, $h_a/h_K=1.0439$, and $r_s=0.8$ gives $\DSP^{3g}=0.966$ in the narrow-width limit. With normalized relativistic spectral functions,
\begin{equation}
 \overline p_{SP}=\int ds\,\widehat\rho_S(s)
 \frac{\lambda^{1/2}(M_X^2,s,m_P^2)}{2M_X},
 \qquad \int ds\,\widehat\rho_S(s)=1,
 \label{eq:S_dsp_fold}
\end{equation}
the central result becomes $0.994$. The range $r_s=0.7$--$1.0$ gives $0.805$--$1.114$, and variation over the enlarged input ranges gives a median of $0.901$ and a 95\% sensitivity interval $0.447$--$1.836$. These quantiles are not confidence intervals.

For reproducibility, the central evaluation uses $\mu_\pi/\mu_K=0.90764$, $h_a/h_K=1.0439$, and $r_s=0.8$. Under the enlarged parameter variation, the current residues are reevaluated for each parameter set from
\begin{equation}
 h_S=m_S\bar f_S,
 \qquad
 \frac{\mu_\pi}{\mu_K}
 =\frac{f_\pi m_\pi^2}{2f_Km_K^2}(1+m_s/m_q).
 \label{eq:S_dsp_residue_relations}
\end{equation}
The varied inputs are $\bar f_a=0.460\pm0.050$ GeV, $\bar f_K=0.445\pm0.050$ GeV, $m_{a_0}=1.474\pm0.019$ GeV, $m_{K_0^*}=1.425\pm0.050$ GeV, and $M_X=2.359\pm0.014$ GeV, with the normal distributions truncated respectively to $0.25$--$0.70$ GeV, $0.25$--$0.70$ GeV, $1.38$--$1.55$ GeV, $1.25$--$1.60$ GeV, and $2.30$--$2.42$ GeV. In addition, $r_s$ is varied according to a normal distribution with mean $0.95$ and standard deviation $0.15$, truncated to $0.50$--$1.40$; $f_\pi=0.1295$--$0.1309$ GeV, $f_K=0.1545$--$0.1570$ GeV, and $m_s/m_q=26.5$--$28.1$ are varied uniformly. A multiplicative finite-width variation of $0.90$--$1.10$ is also included. The quoted sensitivity interval is obtained from $4\times10^5$ parameter sets and quantifies the response to these adopted ranges rather than a statistical uncertainty on $\DSP^{3g}$.

For an exact flavor-singlet hybrid with flavor-blind pair creation,
\begin{equation}
 {\cal A}_H(SP)\propto{\rm Tr}(SP),
\end{equation}
so the three $a_0\pi$ and four $K_0^*K+\cc$ charge amplitudes have the same leading flavor coefficient. Writing the residual spatial overlaps as $F_f$ gives
\begin{equation}
 \boxed{
 \DSP^H=\frac{3p_a}{4p_K}
 \left|\frac{F_{a_0\pi}}{F_{K_0^*K}}\right|^2.}
 \label{eq:S_dsp_hybrid}
\end{equation}
For equal smooth overlaps, $|F_{a_0\pi}/F_{K_0^*K}|=1$, the nominal two-body kinematics gives $\DSP^H=0.872$. The range $\DSP^H\simeq0.85$--$0.87$ quoted in the Letter allows a few-percent smooth variation of the overlap ratio. It is a leading flavor-recoupling estimate and is not included in the matched line-shape comparison below. Reaching the most adverse compact-$4q$ value below would require $|F_{a_0\pi}/F_{K_0^*K}|<0.088$, an additional channel-specific suppression of about one order of magnitude at amplitude level.

\subsection{Physical compact-tetraquark calculation}

The compact tetraquark is calculated with the AL1 and AP1 Hamiltonians of Ref.~\cite{S-Semay:1994ht}. The complete $I=0$, $J^{PC}=0^{-+}$ $P$-wave internal basis contains both $\bar{\bm3}_c\otimes\bm3_c$ and $\bm6_c\otimes\bar{\bm6}_c$ color couplings and the associated spin--flavor configurations. Gaussian Expansion Method wave functions are used, and the variational coefficients satisfy
\begin{equation}
 \sum_j(H_{ij}-EN_{ij})c_j=0.
 \label{eq:S_dsp_GEM}
\end{equation}
The physical seven-dimensional eigenstate is retained after convergence over the $N=6,7,8$ Gaussian range sets. For every annihilation range, the flavor-changing strength is fixed independently from the $\eta$--$\eta'$ spectrum and the tetraquark Hamiltonian is rediagonalized.

For a final channel $f=M_1M_2$, the fall-apart amplitude is
\begin{equation}
 {\cal A}_f=\sum_\alpha c_\alpha
 C_\alpha^{\rm csf}(f)I_\alpha^{\rm sp}(f),
 \label{eq:S_dsp_fallapart}
\end{equation}
where $C_\alpha^{\rm csf}$ is the exact color--spin--flavor recoupling coefficient and $I_\alpha^{\rm sp}$ is the spatial overlap with final meson wave functions generated in the same interaction. Both $a_0(1450)\pi$ and $K_0^*(1430)\bar K+\cc$ are treated as $({}^3P_0)+({}^1S_0)$ pairs in relative $S$ wave with the same fall-apart normalization. No channel-dependent strength is adjusted to the $X(2370)$ data.

The origin of the small $a_0(1450)\pi$ amplitude can be displayed without the common fall-apart normalization. After the spatial components have been summed for each member of the fixed seven-dimensional color--spin--flavor basis, write
\begin{equation}
 {\cal A}_f=\sum_{\alpha=1}^{7}{\cal A}_{\alpha f},
 \qquad
 {\cal C}_f=\frac{|\sum_\alpha{\cal A}_{\alpha f}|^2}
 {\sum_\alpha|{\cal A}_{\alpha f}|^2}.
 \label{eq:S_dsp_coherence_measure}
\end{equation}
Here ${\cal C}_f$ is used only to display interference among the basis components; it is not an additional fitted quantity. At $M_X=2.359$ GeV, $m_{a_0}=1.474$ GeV, and $m_{K_0^*}=1.425$ GeV, the independently calibrated central eigenstates give
\begin{table}[tb]
\caption{Coherent-to-incoherent amplitude measure in the fixed physical seven-dimensional compact-tetraquark basis.}
\label{tab:S_dsp_4q_coherence}
\centering
\begin{tabular}{lcc}
\toprule
Interaction & ${\cal C}_{a_0\pi}$ & ${\cal C}_{K_0^*K}$\\
\midrule
AL1 & $8.69\times10^{-3}$ & $0.230$\\
AP1 & $2.66\times10^{-2}$ & $0.269$\\
\bottomrule
\end{tabular}
\end{table}
The substantially smaller value in the $a_0\pi$ channel shows the destructive recoupling among its color--spin--flavor components. The suppression is present for both interactions before the broad-scalar folding is applied.

With normalized broad-scalar folding, the anomaly scan gives
\begin{align}
 {\rm AL1}:&\quad3.65\times10^{-4}\le\DSP^{4q}
 \le9.91\times10^{-4},\notag\\
 {\rm AP1}:&\quad6.02\times10^{-4}\le\DSP^{4q}
 \le2.13\times10^{-3}.
 \label{eq:S_dsp_4q_scan}
\end{align}
Enlarging the external scalar ranges to $m_{a_0}=1.28$--$1.52$ GeV, $\Gamma_{a_0}=0.12$--$0.40$ GeV, $m_{K_0^*}=1.35$--$1.50$ GeV, and $\Gamma_{K_0^*}=0.16$--$0.40$ GeV raises the largest values only to $2.24\times10^{-3}$ for AL1 and $5.36\times10^{-3}$ for AP1. Extending the parent mass to $2.30$--$2.42$ GeV gives the largest AP1 value $6.17\times10^{-3}$.

\subsection{Matched line-shape test}

The broad $a_0(1450)$ is further described by the two-channel form
\begin{align}
 D_a(s)={}&m_0^2-s-g_{\eta\pi}^2[H_{\eta\pi}(s)-H_{\eta\pi}(m_0^2)]
 -g_{K\bar K}^2[H_{K\bar K}(s)-H_{K\bar K}(m_0^2)]\notag\\
 &+b(s-m_0^2)-i[g_{\eta\pi}^2\rho_{\eta\pi}(s)
 +g_{K\bar K}^2\rho_{K\bar K}(s)],
 \label{eq:S_dsp_CM}
\end{align}
with normalized spectral weight
\begin{equation}
 \widehat w_a(s)=\frac{g_{\eta\pi}^2\rho_{\eta\pi}(s)
 +g_{K\bar K}^2\rho_{K\bar K}(s)}{N_a|D_a(s)|^2},
 \qquad\int ds\,\widehat w_a(s)=1.
\end{equation}
The parameter ranges are $m_0=1.30$--$1.52$ GeV, $\Gamma=0.16$--$0.40$ GeV, $\Gamma_{\eta\pi}/\Gamma_{K\bar K}=0.5$--$2$, and $b=-0.5$--$0.5$, together with the stated $K_0^*$ and parent-mass variations.

For each parameter set $\theta$, the same scalar spectral functions and external kinematics are used in both calculations. The quantity tested is therefore
\begin{equation}
 {\cal S}_{\min}=\min_\theta
 \frac{\DSP^{3g}(\theta)}{\DSP^{4q}(\theta)},
 \label{eq:S_dsp_matched}
\end{equation}
rather than the ratio of unrelated extrema. Across $17\,500$ matched points,
\begin{equation}
 {\cal S}_{\min}=54.98\ ({\rm AL1}),
 \qquad55.97\ ({\rm AP1}).
 \label{eq:S_dsp_separation}
\end{equation}
The adverse values are
\begin{equation}
 \DSP^{3g}\simeq0.363,
 \qquad
 \DSP^{4q}\simeq6.60\times10^{-3}\ ({\rm AL1}),
 \quad6.48\times10^{-3}\ ({\rm AP1}).
 \label{eq:S_dsp_adverse}
\end{equation}
Combining extrema from independent parameter variations would give the looser isolated envelope $\DSP^{4q}<9.52\times10^{-3}$, but it is not used to define the separation because the two extrema do not represent the same line shape and kinematics.

The results entering the Letter are summarized in Table~\ref{tab:S_dsp_summary}. The trigluon-glueball and compact-tetraquark adverse values refer to the matched line-shape comparison, whereas the hybrid value is the leading flavor-recoupling estimate described above.
\begin{table}[tb]
\caption{Summary of the scalar--pseudoscalar ratios calculated in this section. The different entries retain their stated theoretical treatments and should not be interpreted as confidence intervals.}
\label{tab:S_dsp_summary}
\centering
\begin{tabular}{lll}
\toprule
Configuration & $D_{SP}$ & Treatment\\
\midrule
Trigluon glueball & $0.994$ & central finite-width result\\
Trigluon glueball & $0.447$--$1.836$ & enlarged input scan\\
Flavor-singlet hybrid & $0.85$--$0.87$ & smooth-overlap estimate\\
Compact tetraquark & $3.65\times10^{-4}$--$2.13\times10^{-3}$
 & AL1/AP1 anomaly scans\\
Trigluon glueball & $\gtrsim0.363$ & matched line-shape result\\
Compact tetraquark & $\lesssim6.60\times10^{-3}$
 & matched line-shape result\\
\bottomrule
\end{tabular}
\end{table}

\section{The spin-sensitive ratio}
\label{sec:S_rphi}

\subsection{Trigluon-glueball tensor-current contribution}

After one scalar vacuum insertion, the tensor--tensor part of the same Fierz-reduced trigluon current is
\begin{equation}
 -\widetilde J_{3g}\big|_{TT}^{\rm vac}
 =\frac1{12}\epsilon_{\mu\nu\rho\sigma}
 {\rm Tr}(\Sigma T^{\mu\nu}T^{\rho\sigma}).
 \label{eq:S_rphi_TT}
\end{equation}
It is nonzero at leading local order. The tensor-current conventions are
\begin{align}
 \langle0|\bar q\sigma_{\mu\nu}q|V(p,\lambda)\rangle
 &=if_V^T(\epsilon_\mu p_\nu-\epsilon_\nu p_\mu),\notag\\
 \langle0|\bar q\sigma_{\mu\nu}q|b_1(p,\lambda)\rangle
 &=if_{b_1}^T\epsilon_{\mu\nu\alpha\beta}\epsilon^\alpha p^\beta.
 \label{eq:S_rphi_tensor_defs}
\end{align}
The identical-particle factor for $\phi\phi$, the three isoscalar $b_1\rho$ charge combinations, and the polarization sums give
\begin{equation}
 \boxed{
 \RphiBV^{3g}=K_{\phi/BV}
 \left[r_s\frac{(f_\phi^T)^2}{f_{b_1}^Tf_\rho^T}\right]^2,}
 \label{eq:S_rphi_3g}
\end{equation}
with
\begin{equation}
 K_{\phi/BV}=\frac{M_X^2p_\phi^3}
 {3p_b[2X_b^2+m_{b_1}^2m_\rho^2]},
 \qquad
 X_b=\frac{M_X^2-m_{b_1}^2-m_\rho^2}{2}.
 \label{eq:S_rphi_K}
\end{equation}
At $M_X=2.359$ GeV,
\begin{equation}
 p_{\phi\phi}=0.59323~{\rm GeV},
 \qquad p_{b_1\rho}=0.61002~{\rm GeV},
 \qquad K_{\phi/BV}=0.09244.
\end{equation}

All tensor decay constants are evolved to the same renormalization scale before Eq.~(\ref{eq:S_rphi_3g}) is evaluated~\cite{S-Ball:1998sk,S-Jansen:2009yh,S-Braun:2016wnx}. Their common multiplicative evolution cancels in $(f_\phi^T)^2/(f_{b_1}^Tf_\rho^T)$ only after this common-scale conversion. Mutually consistent inputs give
\begin{equation}
 \RphiBV^{3g}=0.08\text{--}0.31,
 \qquad \RphiBV^{3g}\ge0.039
 \quad\hbox{for the conservative factorized range}.
 \label{eq:S_rphi_3g_range}
\end{equation}
Finite-width folding is applied to the broad $b_1\rho$ denominator as
\begin{align}
 \Gamma_{b_1\rho}^{\rm FW}\propto3\int ds_bds_\rho\,
 \widehat\rho_{b_1}(s_b)\widehat\rho_\rho(s_\rho)
 p(M_X;s_b,s_\rho)[2X^2(s_b,s_\rho)+s_bs_\rho],
 \label{eq:S_rphi_bfold}
\end{align}
with normalized spectral functions and the physical boundary. The $16.8$ MeV difference between the nominal recoil momenta also makes the ratio insensitive to a smooth common parent form factor. A genuinely channel-dependent correction is parameterized by
\begin{equation}
 \RphiBV^{3g,{\rm true}}=\RphiBV^{3g,{\rm fact}}|\zeta_{\rm NF}|^2.
\end{equation}
Entering the hybrid range below requires $|\zeta_{\rm NF}|<\sqrt{0.006/0.039}=0.392$.

\subsection{Hybrid spin recoupling and relativistic lifting}

For the lowest pseudoscalar hybrid, $S_{q\bar q}=1$. In a non-flip spin-triplet decay operator the vector--vector amplitude contains
\begin{equation}
 \left\{\begin{matrix}
 \tfrac12&\tfrac12&1\\
 \tfrac12&\tfrac12&1\\
 1&1&1
 \end{matrix}\right\}=0,
 \label{eq:S_rphi_9jzero}
\end{equation}
whereas the spin-singlet/vector recoupling relevant to the denominator is
\begin{equation}
 \left\{\begin{matrix}
 \tfrac12&\tfrac12&1\\
 \tfrac12&\tfrac12&1\\
 1&0&1
 \end{matrix}\right\}=\frac{\sqrt6}{18}\ne0
 \label{eq:S_rphi_9jallowed}
\end{equation}
\cite{S-Page:1998gz,S-Farina:2023oqk}. Equation~(\ref{eq:S_rphi_9jzero}) is a leading spin-recoupling zero, not an exact QCD selection rule.

The correction is evaluated with the complete pair-creation vertex
\begin{equation}
 {\cal V}=\bar u(\bm p_3)\bm\gamma\cdot\bm\epsilon_gv(\bm p_4).
\end{equation}
Expanding the Dirac spinors gives
\begin{align}
 {\cal V}\propto\chi_3^\dagger\left[
 \bm\sigma\cdot\bm\epsilon_g+
 \frac{(\bm\sigma\cdot\bm p_3)(\bm\sigma\cdot\bm\epsilon_g)
 (\bm\sigma\cdot\bm p_4)}{(E_3+m_3)(E_4+m_4)}
 \right]\eta_4.
 \label{eq:S_rphi_Dirac}
\end{align}
The first term and the spin-triplet part of the second term retain the zero in Eq.~(\ref{eq:S_rphi_9jzero}). The first nonvanishing spin-singlet piece is
\begin{equation}
 C_0(\bm p_3,\bm p_4)=
 \frac{\bm\epsilon_g\cdot(\bm p_3\times\bm p_4)}
 {(E_3+m_3)(E_4+m_4)},
 \label{eq:S_rphi_C0}
\end{equation}
which is of order $v^2$ because it requires the lower component of both created spinors.

The momentum dependence is retained in the full overlap
\begin{equation}
 {\cal I}_f=\int\frac{d^3\bm k}{(2\pi)^3}
 \frac{d^3\bm r}{(2\pi)^3}
 \Psi_H(\bm k,\bm r)\Phi_A^*(\bm q_A)\Phi_B^*(\bm q_B)
 {\cal K}_f(\bm k,\bm r;\bm p_f).
 \label{eq:S_rphi_6D}
\end{equation}
The common coupling, color factor, and parent normalization are canceled before the ratio is formed. The converged light-flavor calculation gives
\begin{equation}
 \frac{\Gamma_H(\rho\rho)}{\Gamma_H(b_1\rho)}
 =8.28(5)\times10^{-3}
\end{equation}
at narrow width and approximately $9.9\times10^{-3}$ after folding. Using the same wave functions and kernel gives
\begin{equation}
 \frac{\Gamma_H(\phi\phi)}{\Gamma_H(\rho\rho)}=0.179,
 \qquad
 \frac{I_\phi}{I_\rho}=0.898.
\end{equation}
Consequently,
\begin{equation}
 \RphiBV^{H}=1.48\times10^{-3}
 \end{equation}
at narrow width and about $2.0\times10^{-3}$ after spectral folding. Variations of the constituent masses, wave-function widths, strange/light conversion, daughter line shapes, and parent kinematics remain below
\begin{equation}
 \RphiBV^H<0.006.
 \label{eq:S_rphi_H_range}
\end{equation}
The conservative separation from the trigluon-glueball factorized edge is therefore $0.039/0.006=6.5$.

\section{Combined interpretation and experimental extraction}
\label{sec:S_combined}

The most adverse values obtained for the unmixed assignments are
\begin{align}
 \DSP^{4q}&\lesssim6.6\times10^{-3},
 &\DSP^{3g}&\gtrsim0.363,
 &\DSP^H&\simeq0.85\text{--}0.87,\notag\\
 \RphiBV^H&<0.006,
 &\RphiBV^{3g}&\gtrsim0.039.
 \label{eq:S_combined_regions}
\end{align}
The interpretation boundaries used in the Letter are deliberately placed inside the gaps,
\begin{equation}
 0.02<\DSP<0.20,
 \qquad
 0.01<\RphiBV<0.03.
 \label{eq:S_combined_buffers}
\end{equation}
They are not likelihood intervals or confidence limits. A measurement inside either interval is not an overlap of the calculated pure-state regions; it indicates that configuration mixing, additional decay dynamics, or a revision of one of the assumed assignments must be examined. For a mixed state,
\begin{equation}
 |X\rangle=\sum_\alpha c_\alpha|\alpha\rangle,
 \qquad
 {\cal M}_f=\sum_\alpha c_\alpha{\cal M}^{(\alpha)}_f,
 \end{equation}
so a linear interpolation between pure-state width ratios is generally incorrect.

The common radiative-production factor cancels from both observables:
\begin{equation}
 \frac{{\cal B}(J/\psi\to\gamma X\to\gamma f_i)}
 {{\cal B}(J/\psi\to\gamma X\to\gamma f_j)}
 =\frac{\Gamma(X\to f_i)}{\Gamma(X\to f_j)},
 \label{eq:S_combined_cancel}
\end{equation}
after all daughter branching fractions have been unfolded. The charge-complete parent modes are
\begin{align}
 a_0\pi:&\quad a_0^+\pi^-,\ a_0^-\pi^+,\ a_0^0\pi^0,\notag\\
 K_0^*K+\cc:&\quad K_0^{*+}K^-,\ K_0^{*0}\bar K^0,
 \ K_0^{*-}K^+,\ \bar K_0^{*0}K^0,\notag\\
 b_1\rho:&\quad b_1^+\rho^-,\ b_1^-\rho^+,\ b_1^0\rho^0.
 \label{eq:S_combined_charges}
\end{align}
The relative phases inside each $I=0$ combination are fixed by the chosen isospin convention and should be imposed coherently. The $\phi\phi$ amplitude must be symmetrized, and its theoretical width already contains the identical-particle factor.

For phase-space coordinates $\Phi$, a coherent amplitude model has
\begin{equation}
 {\cal A}_{\rm tot}(\Phi)=\sum_\alpha c_\alpha A_\alpha(\Phi),
 \qquad
 \frac{dN}{d\Phi}\propto\epsilon(\Phi)|{\cal A}_{\rm tot}(\Phi)|^2.
 \label{eq:S_combined_event}
\end{equation}
The acceptance-corrected diagonal intensity of a specified quasi-two-body amplitude is
\begin{equation}
 {\cal I}_i=\int d\Phi\,|c_iA_i(\Phi)|^2,
 \label{eq:S_combined_intensity}
\end{equation}
while the coherent fit also retains
\begin{equation}
 {\cal I}_{ij}^{\rm int}=2{\rm Re}\int d\Phi\,
 c_ic_j^*A_i(\Phi)A_j^*(\Phi).
\end{equation}
Because the fit fractions are nonadditive in the presence of these terms, raw fractions from separate amplitude models must not be divided. If $B_i^{\rm dau}$ is the product of daughter branching fractions, the two parent ratios are
\begin{equation}
 \DSP=\frac{{\cal I}_{a_0\pi}/B_{a_0\pi}^{\rm dau}}
 {{\cal I}_{K_0^*K}/B_{K_0^*K}^{\rm dau}},
 \qquad
 \RphiBV=\frac{{\cal I}_{\phi\phi}/B_{\phi\phi}^{\rm dau}}
 {{\cal I}_{b_1\rho}/B_{b_1\rho}^{\rm dau}}.
 \label{eq:S_combined_extract}
\end{equation}

The experimental results should be reported through a covariance matrix or profile likelihood for
\begin{equation}
 \bm z=(\ln\DSP,\ln\RphiBV)^T,
\end{equation}
with common detector and reconstruction systematics treated as correlated nuisance parameters. The $X(2370)$ pole parameters should be shared across the relevant final states. The scalar-line-shape, nonresonant, and neighboring-wave variations should be included for $D_{SP}$; the broad $b_1$ and $\rho$ line shapes, multipion-wave covariance, and four-kaon backgrounds should be included for $R_{\phi/BV}$. Experimentally, $D_{SP}$ should be determined first. If its experimental range lies below $0.02$, the result is compact-$4q$-like within the stated set; if it lies above $0.20$, $R_{\phi/BV}$ is then used to distinguish the hybrid-like region below $0.01$ from the trigluon-glueball-like region above $0.03$.

\makeatletter
\let\@FMN@list\@empty
\makeatother

\end{document}